\documentclass[%
 reprint,
superscriptaddress,
nofootinbib,
bibnotes,
 amsmath,amssymb,
]{revtex4-1}
\usepackage{fontenc}
\usepackage{inputenc}
\usepackage{booktabs}
\usepackage{wasysym}
\usepackage{xcolor}
\usepackage{gensymb}
\usepackage{amssymb}
\usepackage[english]{babel}
\usepackage{amsmath}
\usepackage{tensor}
\usepackage{graphicx}
\usepackage{dcolumn}
\usepackage{bm}

 \newcommand{\RNum}[1]{\uppercase\expandafter{\romannumeral #1\relax}}
\begin{document}

\preprint{APS/123-QED}

\title{Modified Gravitational Waves Across Galaxies from Macroscopic Gravity}

\author{Giovanni Montani}
\altaffiliation{giovanni.montani@enea.it}

\affiliation{
	ENEA, Fusion and Nuclear Safety Department, C. R. Frascati,
	Via E. Fermi 45, 00044 Frascati (Roma), Italy\\
}%
\affiliation{
	Physics Department, “Sapienza” University of Rome,
	P.le Aldo Moro 5, 00185 Roma, Italy
}%
 
\author{Fabio Moretti}%
\email{fabio.moretti@uniroma1.it}

\affiliation{
	Physics Department, “Sapienza” University of Rome,
	P.le Aldo Moro 5, 00185 Roma, Italy
}%




\date{\today}

\begin{abstract}
\noindent We analyze the propagation of gravitational 
waves in a medium containing bounded 
subsystems ("molecules''), able to induce significant Macroscopic Gravity effects.

\noindent We establish a precise constitutive relation between the 
average quadrupole and the amplitudes of a vacuum gravitational wave, via the geodesic deviation equation. Then we determine
the modified equation for the wave inside the medium and the associated dispersion relation. 

\noindent A phenomenological analysis shows that 
anomalous polarizations of the wave emerge 
with an appreciable experimental detectability 
if the medium is identified with a typical galaxy. 

\noindent Both the modified dispersion relation 
(wave velocity less than the speed of light) 
and anomalous oscillations modes could be detectable by the incoming LISA or pulsar timing arrays experiments, having the appropriate size to see the concerned wavelengths
(larger than the molecular size) and 
the appropriate sensitivity to detect the 
expected deviation from vacuum General Relativity. 
 
\end{abstract}

\pacs{Valid PACS appear here}
\maketitle


\section{Introduction}
The Einsteinian theory of gravity offers 
a predictive tool to investigate the Universe structure on very different
spatial scales, from the Hubble flow to 
the Solar system morphology \cite{1,2,3,4}. 

However, Einstein field equation in matter 
is approached by considering the space-time metric and the physical sources
as 
continuous fields, having a differentiable 
(at least of class $C^2$) profile. 
Nonetheless, as it turns out looking at 
the real morphology of astrophysical systems \cite{5,6,7,8}, 
this notion of continuum is valid only 
on an average sense. In fact, the 
matter sources are typically characterized by a discrete nature, for instance,
stars or galaxies are merely point-like sources, when treated on a scale much larger than their typical size. 
As a consequence, also the space-time geometry and the associated metric
tensor 
acquire a discrete nature: clumpiness
of the sources induces irregularities 
in the Einsteinian manifold. 

Therefore, an appropriate treatment of 
the implementation of Einstein equation to real astrophysical systems
requires a suitable procedure for 
averaging both matter and geometry. 
It is easy to realize how the definition of an averaging procedure of the
space-time be a non-trivial task, mainly due to the non-linearity of the
gravitational interaction: the average of the Einstein tensor is not the
Einstein tensor in the averaged 
metric, but a complicate set of correlation functions comes out \cite{9,10,11}. 

Another subtle question, strictly connected 
with the above considerations, is the 
existence of bounded subsystems within a 
matter medium, for instance
the presence of binary systems and open 
clusters within the galaxy \cite{12}. 
Such subsystems behave as real "gravitational molecules'' and when the
gravitational field interact with them, 
their structure is altered with a consequent 
gravitational backreaction. 
Thus, we see how it is, in general, necessary to deal with "Macroscopic
Gravity'' 
physics, in close analogy to what happens 
in the case of the electromagnetism within matter \cite{13,14,15}. For relevant analyses
of 
macroscopic gravity, see \cite{16,17}, where 
covariance requirements and constitutive 
relations are addressed. In particular in \cite{16}
, 
the theory is constructed in close analogy to electromagnetism, taking the
Weyl tensor as the gravitational 
counterpart of the electromagnetic tensor field. 
Furthermore, an interesting technique 
to separate the source energy momentum 
tensor is provided, reconstructing the continuous matter field plus a
quadrupole term, associated to the molecular structure. 
In \cite{17}, the case of high frequency 
gravitational waves is considered 
on a generic background and a closed 
set of constitutive equations is fixed, 
relating the quadrupole term to the background curvature. 

Several authors have dealt with the problem of gravitational waves in a matter medium: in \cite{18,19,20} is analyzed the amplitude damping arising from the propagation in a dissipative fluid, characterized by a definite viscosity and in \cite{21} constraints on the viscosity of the Universe given by observations of gravitational waves are addressed; in \cite{22,23,24,25,26} is studied the modification in the dispersion relation when the matter medium is a collisionless kinetic gas, whose density perturbation are governed by the Vlasov equation; with the same method is studied the interaction of cosmological gravitational waves with neutrinos \cite{lamo,labemo,belamo} and spinning particles \cite{milamo}; a review of results mainly coming from the kinetic approach can be find in \cite{thorne}, where is stated that the effects should be in any case negligible; the possibility of Landau damping of gravitational waves traveling in a Robertson-Walker zero curvature spacetime filled with a perfect fluid is considered in \cite{27}, whereas in \cite{28} a modified dispersion relation and extra modes of polarizations are showed to appear when the medium considered is a spherical cloud (perfect fluid in Schwarzschild metric); the case of pressureless matter (dust) is studied in \cite{29}, where is outlined the appearance of extra modes, and in \cite{30}, where is investigated the possibility of amplitude damping.

Here, we analyze the problem of the 
macroscopic gravity theory, as referred to 
the propagation of gravitational waves in 
a molecular medium. We retain the 
splitting, proposed in \cite{16}, between the free continuum energy-momentum tensor
and 
the quadrupole term, then we calculate the constitutive relation, \textit{i.e.} the
form 
of the molecule quadrupole from the geodesic deviation equation, by perturbing the molecule structure with the physical degrees of freedom of the wave, namely the plus and cross polarizations. We construct the
quadrupole tensor via the geodesic deviation vector and then express it in
terms of 
the vacuum gravitational wave amplitudes. 
As a result, we get a closed wave equation 
in the medium, which provides us with five 
effective physical degrees of freedom. 
We characterize this modified scenario 
for the weak gravitational field, by 
studying the wave polarizations and the dispersion relation descending from the macroscopic field equation, then considering some specific example 
of molecular structures. 
Our study outlines a dispersion profile 
of a gravitational wave in a molecular medium and new modes of oscillation to
be searched in the experimental devices. 

This work can be summarized as follows: in section \RNum{2} a constitutive equation relating the components of the quadrupole tensor to the amplitudes of an incoming gravitational wave will be derived via the geodesic deviation equation, taken in the co-moving frame; in section \RNum{3} we will calculate the wave equation descending from the adoption of the constitutive relation derived in section \RNum{2} and we will analyze the phenomenological features implied, emphasizing the discrepancies with the standard case; in section \RNum{4} we will compare our study to the previous literature, stressing the conceptual and technical differences of our approach;  in section \RNum{5} we will establish two different models of macroscopic medium, giving some quantitative estimates of the observable effects; in section \RNum{6} we will comment the results obtained.
\section{Quadrupole tensor and constitutive relation}
 This work is included in the theoretical framework of linearized gravity. As usual in this context, the metric $g_{\mu\nu}$ can be written as a sum of Minkowksi flat metric $\eta_{\mu\nu}=diag \left(-1,1,1,1\right)$ plus a perturbation $h_{\mu\nu}$, small enough to be neglected the contribution of terms of quadratic order. 
 
  The approach to Macroscopic Gravity proposed in \cite{16} is based on the hypothesis that the material medium can be described as a set of point-like masses grouped into molecules. 
  This means that, defining $z_i^\mu$ as the world line of the $i$-th particle and $y_a^\mu$ as the world line of the center of mass of the $a$-th molecule, the four-vector $s_i^\mu=z_i^\mu-y_a^\mu$ satisfies 
  
  \begin{equation}
  |s_i^\mu|\ll D,
  \end{equation}
  where $D$ indicates the mean distance between the centers of mass of the molecules.\footnote{If we assume the medium to be homogeneous and isotropic the quantity $D$ is proportional to the density of molecules $N$, as $D\propto N^{-\frac{1}{3}}$.} It must be stressed that $s_i^\mu=z_i^\mu\left(\tau_i\right)-y_a^\mu\left(\tau_a\right )$, being $\tau_i$ and $\tau_a$ the proper times of the $i$-th particle and of the $a$-th molecule respectively. This means that $s_i^\mu$ depends both on $\tau_i$ and $\tau_a$, so a relation between the two proper times must be fixed. This is done via the equation 
  	\begin{equation}\label{uaua}
  \eta_{\mu\nu}	s_i^\mu u_a^\nu=0,
  	\end{equation} 
being $u_a^\mu$ the velocity of the center of mass of the $a$-th molecule, \textit{i.e.} $u_a^\mu=d y_a^\mu/d\tau_a$. This equation states that $s_i^\mu$ is a spacelike vector and reduces to $s_i^0=0$ in the comoving frame. 
As shown in \cite{16} a stress energy tensor describing a set of point-like masses can be written, after the application of Kaufman's molecular moments method \cite{14} (denoted as $\left < \cdot \right >$) and expanding all the involved quantities up to the second order in $s_i^\mu$, as
\begin{equation}\label{tmedio}
\left < T_{\mu\nu} \right > = T_{\mu\nu}^{(f)}+\dfrac{c^2}{2}{Q}_{\mu\rho\nu\sigma ,}^{\quad \; \; \; \, \rho \sigma},
\end{equation}
where $T_{\mu\nu}^{(f)}$ is the stress energy tensor that describes a set of free particles, \textit{i.e.} the centers of mass of the molecules, and ${Q}_{\mu\rho\nu\sigma}$ is the quadrupole polarization tensor, whose expression in terms of $s_i^\mu$ and $\dot{s}_i^\mu$ can be found in \cite{16}. The notation $V,_{\mu}$ indicates $\dfrac{\partial}{\partial x^\mu}V$ while the dot is intended as a derivation with respect to the proper time.
The quadrupole tensor shares the same set of symmetries of the Riemann, \textit{i.e.} is characterized by twenty free components: these twenty quantities describe the structure of the molecule. 

Our aim is to calculate a linear constitutive relation between the quadrupole polarization tensor of a spherical molecule and the amplitudes of an incoming gravitational wave. The presence of the wave alters the molecular structure and the resulting variation of the components of the quadrupole tensor can be expressed in terms of the amplitudes of the wave itself. The problem will be treated on a linear level, \textit{i.e.} neglecting the changes in the metric coming from the alteration in the background quadrupole. We will follow an approach analogous to the electromagnetic case, replacing Newton's law of motion with the geodesic deviation equation
\begin{equation}  \label{devgeod}
\dfrac{D^2\xi^\mu}{d\tau^2}=R^\mu_{\;\, \nu\rho\sigma}u^\nu u^\rho \xi^\sigma,
\end{equation}
where $u^\mu=dx^\mu/d\tau$ is the velocity of the observer, $\tau$ is the proper time measured by the observer and $\xi^\mu$ is the infinitesimal vector connecting adjacent geodesics: in the following calculation it will be identified with $s_i^\mu$. We stress the fact that the geodesic deviation equation holds for infinitesimal vectors, \textit{i.e.} vectors that are small when compared with the typical scale of variation of the gravitational field. If we define the macroscopic parameter $L$, the radius of the molecule, as the average performed on all the particles belonging to that molecule of the quantities $|s_i^\mu|$, we have that the geodesic deviation equation is valid if 
	\begin{equation} \label{conditio}
	\lambdabar \gg L,
	\end{equation}
being $\lambdabar$ the reduced wavelength of the gravitational radiation. We write the geodesic deviation equation in a frame comoving with the center of mass of the molecule, setting $u^\mu=c\left(1,0,0,0 \right)$. It follows, from \eqref{uaua}, that $\xi^0=0$: it can be shown that, under this circumstance, the covariant derivative $D^2/d\tau^2$ coincides with the ordinary derivative $d^2/d\tau^2$ \cite{oharuf}. In addition to this we have that, in the linear approximation, the coordinate time $t$ is equal to the proper time $\tau$. In computing the components $R^\mu_{\;\, 00\sigma}$ of the Riemann tensor, one has to take into account that the latter is invariant in linearized gravity, rather than covariant as in the full theory: we can calculate $R^\mu_{\;\, 00\sigma}$ in a convenient frame, such as the $TT$ frame, where the form of the metric is simple.
The external field that we use to perturb the molecule is a purely plus polarized vacuum gravitational wave whose direction of propagation is coincident with the $z$ axis: this means that the only relevant dynamic component of the wave, \textit{i.e.} the plus polarization, can be expressed in terms of plane waves $e^{i(\omega(k)t-kz)}$, where $\omega(k)=ck$ and $k$ is the wavenumber.  \\
This being said, we calculate the components $R^\mu_{\;\, 00\sigma}$ from the metric 
\begin{equation}\label{matrice2}
h_{\mu \nu}= \left(\begin{matrix}-\frac{2\phi}{c^2} & 0 & 0 & 0  \\ 0& -\frac{2\phi}{c^2}+a & 0 & 0 \\ 0 &0&-\frac{2\phi}{c^2}-a & 0\\0 & 0 & 0 & -\frac{2\phi}{c^2} \end{matrix}\right),
\end{equation}
where $\phi$ is the static Newtonian potential generated by the molecule itself and $a$ is the plus polarization.\\
The geodesic deviation equation (\ref{devgeod}) takes the form
\begin{equation} \begin{split}
\dfrac{1}{c^2}\dfrac{d^2\xi^i}{dt ^2}&=\frac{1}{2} \eta^{ij} \left ( h_{jk,00} + h_{00,jk} \right)\xi^k=\\
&=\frac{1}{2} \eta^{ij} \left ( h_{jk,00} -\frac{2}{c^2}\phi,_{jk} \right)\xi^k.
\end{split}\end{equation}The Newtonian potential inside a spherical distribution of mass, taking the mass-energy density $\rho(x)$ to be a constant $\rho_0$, is
\begin{equation}
\phi(r)=\frac{2}{3} \pi G \rho_0 r^2
\end{equation}
where $r$ is the distance from the center of the sphere: then inside the molecule the tensor of the second derivatives of the potential $\eta^{ij}\phi,_{jk}$ can be written\footnote{Despite the fact that the potential $\phi\propto r^2$ that we use in this calculation is not the potential associated with a large number of particles, we outline the fact that the molecule is imagined as a bounded and stable system: the particles can be assumed to be confined in a region sufficiently close to the center, so that their motion can be described as small oscillations around a parabolic minimum.}
\begin{equation}\label{matrice1}
\eta^{ij}\phi,_{jk}=\omega_0^2 \delta^i_{\;j},
\end{equation}
where $\omega_0^2=\frac{4}{3} \pi G \rho_0$. \\
The second time derivatives of the spatial components of the wave can be calculated as
\begin{equation}
h_{ij,00}=-\dfrac{\omega^2(k)}{c^2}\; 
\left(\begin{matrix} a & 0 & 0  \\0&-a & 0 \\ 0 & 0 &0 \end{matrix}\right).
\end{equation}
We find the following system of differential equations
\begin{equation} \label{syst}
\left(\begin{matrix}\ddot{\xi}^1 \\\ddot{\xi}^2\\ \ddot{\xi}^3 \end{matrix}\right ) =-\omega_0^2\left(\begin{matrix} 1+\epsilon & 0 & 0 \\0&1-\epsilon & 0\\ 0 & 0 & 1 \end{matrix}\right)\left(\begin{matrix}\xi^1 \\\xi^2\\ \xi^3 \end{matrix}\right)
\end{equation}
where we have defined the parameter $\epsilon=\dfrac{\omega^2(k)}{2\omega_0^2}a$. If we assume that $\omega \ll \omega_0$, the parameter $\epsilon$ results to be small with respect to the unity. 
Moreover, this assumption implies that $a$ changes with a typical time much greater than the time of travel of the wave inside the molecule: in other words, when we integrate the system \eqref{syst} we can consider $\epsilon$ as a constant. We obtain
\begin{equation}\label{systpert}
\begin{split}
\xi^1(t)&=\alpha_1 \cos\left(\omega_0\sqrt{1+\epsilon}t+\beta_1 \right ) \\
\xi^2(t)&=\alpha_2 \cos\left(\omega_0\sqrt{1-\epsilon}t+\beta_2 \right ) \\
\xi^3(t)&=\alpha_3 \cos\left(\omega_0t+\beta_3 \right )
\end{split}
\end{equation}
where $\alpha_1$, $\alpha_2$, $\alpha_3$, $\beta_1$, $\beta_2$ and $\beta_3$ are constants of integration.
This allows us to calculate the components of the quadrupole tensor from its expression, as in \cite{16}:
\begin{equation}\label{quadrupol}
Q_{i0j0}=N \left < \int \rho s_i s_j d^3x \right >=\frac{1}{2}MN \left<\alpha_i ^2\right >  \delta_{ij}
\end{equation} 
where $M$ is the total mass of the molecule and $N$ the density of molecules. \\
In the absence of the gravitational wave one would have found
\begin{equation}\label{systimpert}
\begin{split}
\xi^1(t)&=A_1 \cos\left(\omega_0t+c_1 \right ) \\
\xi^2(t)&=A_2 \cos\left(\omega_0t+c_2 \right ) \\
\xi^3(t)&=A_3 \cos\left(\omega_0t+c_3 \right ),
\end{split}
\end{equation}
implying the following expression for the quadrupole tensor
\begin{equation}\label{quadrupintrinseco}
\left(Q_{i0j0}\right)_0=\frac{1}{2}MN L ^2 \delta_{ij},
\end{equation} 
having defined 
\begin{equation}\label{ellequadro}
L^2= \left <A_1^2 \right >=\left <A_2^2 \right >=\left <A_3 ^2\right >
\end{equation}
the square of the typical radius of the molecule, since the spatial oscillation are, on average, comparable with the size of the molecule. Specifically, the definition \eqref{ellequadro} is allowed by the assumption of homogeneity and isotropy of the molecule. If we average the amplitudes of the trajectories of the particles we get the same value along all the three directions: in this sense the molecule is (averagely) spherical.  \\ 
If we consider a harmonic oscillator with equation of motion
\begin{equation}
x(t)=A \cos \left( \omega_0 t + c \right )
\end{equation}
and we imagine to change the frequency from $\omega_0$ to $\omega'$ when $x=x_0$, after which
\begin{equation}
x(t)=\alpha \cos \left( \omega' t + c' \right )
\end{equation}
we have, on equating kinetic energies\footnote{In calculating, through formula (22), the modified amplitudes of the particles' trajectory, one has to perform an average over a time interval $\Delta T$ comparable with the period of motion of the point particles. Since $\omega\ll\omega_0$, we can assume that the gravitational wave is seen by the point particles as a constant at any time during the averaging. In addition to this, the turning on of the wave is considered as an instantaneous process, in the sense that it happens on a time scale much smaller than the period on which the average is taken.} when $x=x_0$, as firstly done in \cite{16}, that
\begin{equation}
\alpha^2=A^2+\dfrac{\omega_0^2-\omega'^2}{\omega'^2}\left( A^2-x_0^2 \right).
\end{equation}
If we take the mean value of this equation we get
\begin{equation} \label{blabla}
\left< \alpha_i^2\right>=L^2 +\dfrac{\omega_0^2-\omega_i^2}{\omega_i^2}\left( L^2-\left<x_0^2\right> \right),
\end{equation} 
having identified $\alpha$ with one of the amplitudes $\alpha_i$ in \eqref{systpert}, $\omega'$ with $\omega_i$, the modified angular frequencies in \eqref{systpert}, and $A$ with one of the amplitudes $A_i$ in \eqref{systimpert}.
In order to calculate the term $\left<x_0^2\right> $ we have to remember that $x_0$ is a random point of the unperturbed trajectory, so we can take one of the equations of motion in \eqref{systimpert} and perform an average over time:
\begin{equation}
\left<x_0^2\right>=\dfrac{1}{2}\left<A_i^2\right>=\dfrac{1}{2}L^2,
\end{equation}
where the last equality is implied by \eqref{ellequadro} and the factor $\frac{1}{2}$ comes from the mean value of the square of the cosine. In this way equation \eqref{blabla} can be recast in
\begin{equation}
\left<\alpha_i^2 \right >  =\dfrac{\omega_0^2+\omega_i^2}{2\omega_i^2}L^2.
\end{equation}
Hence, at first order in $\epsilon$, we get
\begin{equation}\begin{split}\label{ampiezze}
\left<\alpha_1 ^2\right >&= \dfrac{\omega_0^2 (1+\epsilon)+\omega_0^2}{2\omega_0^2(1+\epsilon)}L^2 \simeq \left ( 1-\frac{\epsilon}{2}\right)L^2 \\
\left<\alpha_2 ^2\right >  &= \dfrac{\omega_0^2 (1-\epsilon)+\omega_0^2}{2\omega_0^2(1-\epsilon)} L^2\simeq  \left ( 1+\frac{\epsilon}{2}\right)L^2 \\
\left<\alpha_3 ^2 \right >  &=L^2.
\end{split}\end{equation}
Casting what we have just found into equation (\ref{quadrupol}) yields to
\begin{equation}\label{quadru}
Q_{i0j0}=\dfrac{MNL^2}{2}\left(\begin{matrix} 1-\dfrac{\epsilon}{2} & 0 & 0  \\0&1+\dfrac{\epsilon}{2} & 0 \\ 0 & 0 & 1 \end{matrix}\right).
\end{equation}
In order to enounce a constitutive relation we have to express this tensor in terms of the degrees of freedom perturbing the metric.
By comparing equation (\ref{quadru}) with the expressions of $h_{ij}$ and $\phi,_{ij}$, (\ref{matrice2}) and (\ref{matrice1}), we get:
\begin{equation}\label{matrela1}
Q_{i0j0}=\epsilon_g \left(\frac{2}{c^2}\phi,_{ij}+\frac{1}{2}h_{ij,00} \right), 
\end{equation}
where the gravitational dielectric constant is 
\begin{equation}\label{epsilone}
\epsilon_g=\dfrac{MNL^2c^2}{4\omega_0^2}.
\end{equation}
With the same method we can calculate the other components of the quadrupole tensor: this results in
\begin{equation}\label{matrel2}
\begin{split}
Q^{0ijk}&=0\\
Q^{ijkl} &\simeq 0.
\end{split}
\end{equation}
The components $Q^{ijkl}$ result proportional to combinations of the components of the wave through another dielectric constant $\epsilon'_g$ that is small compared to $\epsilon_g$:
\begin{equation}
\dfrac{\epsilon'_g}{\epsilon_g}=\dfrac{\omega_0^2L^2}{3c^2}=O\left( \beta^2 \right),
\end{equation}
so we will ignore their contribution to the field equation.
A completely analogous calculation can be performed by perturbing the molecule with a purely cross polarized gravitational wave. It can be shown that the material relation has the same form as in \eqref{matrela1} and \eqref{matrel2}. This is certainly not a surprise, given the spheric symmetry of the molecule and considering that the effect of the cross polarization is the same of the plus polarization rotated by an angle of $45\degree$.

\section{Field equation and phenomenology}
We will now write the field equation of Macroscopic Gravity
\begin{equation}\label{FE}
 \left < G_{\mu\nu}\right>=\chi \left ( T_{\mu\nu}^{(f)}+\dfrac{c^2}{2}{Q}_{\mu\rho\nu\sigma ,}^{\quad \; \; \; \, \rho \sigma}\right) \quad , \quad \chi=\dfrac{8\pi G}{c^4}
\end{equation} 
that governs the dynamics of the average gravitational field within the macroscopic medium, making use of the set of constitutive relations (\ref{matrela1}) and (\ref{matrel2}). We recall that the macroscopic field dynamics lives on a spatial scale much greater than the molecular separation $D$.
Noticing that the set of centers of mass of the molecules behave as a dust, \textit{i.e.} pressure can be neglected, the free stress-energy tensor is written
\begin{equation}
T^{(f)}_{00}=c^2\rho \qquad , \quad T^{(f)}_{0i}=0 \quad , \quad T^{(f)}_{ij}=0,
\end{equation}
where $\rho$ is the smoothed out mass density of the centers of mass of the molecules.
The left handed side of the field equation \eqref{FE} is the averaged Einstein tensor, expressed up to the first order in $\left <\bar{h}_{\mu\nu}\right >=\left <h_{\mu\nu}\right >-\frac{1}{2}\eta_{\mu\nu}\left <h\right >$, being $\left <h\right >$ the trace of the averaged metric $\left <h_{\mu\nu}\right >$: the tensorial nature of the field equation allows us a gauge freedom connected with the diffeomorphism invariance of the theory. We exploit this freedom by imposing the usual Hilbert gauge fixing 
\begin{equation}\label{hilb}
\partial^\mu \left<\bar{h}_{\mu\nu}\right >=0.
\end{equation}
With this choice Einstein tensor reduces to 
\begin{equation}
\left <G_{\mu\nu}\right >=-\dfrac{1}{2}\Box \left <\bar{h}_{\mu\nu}\right >.
\end{equation}
We stress the fact that, after the imposition of the gauge fixing \eqref{hilb}, one can express four components of the wave in terms of the six remaining: the number of degrees of freedom of the wave decreases from ten to six. At this level, if we impose that the direction of propagation of the wave is coincident with the $z$ axis, we can represent $\left <\bar{h}_{\mu\nu}\right >$\footnote{From this time forth we drop the symbol $\left < \cdot \right >$ for the sake of convenience.} as
	\begin{equation}\label{metricamezzo}
\bar{h}_{\mu\nu}=\left(\begin{matrix} -\frac{4 \phi}{c^2}+\left( \frac{ck}{\omega}\right)^2\bar{h}_{33}& -\frac{ck}{\omega}\bar{h}_{13}  & -\frac{ck}{\omega}\bar{h}_{23} & -\frac{ck}{\omega}\bar{h}_{33} \\\vdots&\bar{h}_{11} & \bar{h}_{12}&\bar{h}_{13} \\ \vdots & \cdots & \bar{h}_{22}&\bar{h}_{23} \\ \vdots &\cdots &\cdots & \bar{h}_{33}\end{matrix}\right).
\end{equation}
We express the constitutive relations \eqref{matrela1}, \eqref{matrel2} in terms of $\bar{h}_{\mu\nu}$ and we compute the components of the field equation \eqref{FE}.
The $(00)$ component results in 
\begin{multline}\label{eq00}
\bigtriangleup \phi -\dfrac{c^4k^2}{4\omega^2}\Box \bar{h}_{33} = 4 \pi G \rho+ \dfrac{4 \pi G \epsilon_g}{c^2}\bigtriangleup^2 \phi + \\
+\pi G \epsilon_g \dfrac{c^2k^2}{\omega^2} \partial_0^4 \bar{h}_{33}-\dfrac{\pi G \epsilon_g}{2}\bigtriangleup\partial_0^2 \bar{h},
\end{multline}
being $\bigtriangleup$ Laplace operator, $\Box=-\partial_0^2+\bigtriangleup$ the d'Alembertian and $\bar{h}$ the trace of the \textit{radiative} part of the metric $\bar{h}_{\mu\nu}$\footnote{The static contribution to the trace coming from the Newtonian potential is negligible in this context, due to the fact that $\bar{h}$ appears in the field equations only through its time derivatives.}, namely 
\begin{equation}\label{trace}
\bar{h}=\bar{h}_{11}+\bar{h}_{22}+\left(1-\dfrac{c^2k^2}{\omega^2}\right)\bar{h}_{33}.
\end{equation}
Given the fact that in \eqref{eq00} the variables $\bar{h}_{33}$ and $\bar{h}$ are functions of time, whilst $\phi$ is static, we argue that the following equations must hold, separately:
\begin{equation}\label{newt}
\bigtriangleup \phi=4 \pi G \rho+\dfrac{4 \pi G \epsilon_g}{c^2}\bigtriangleup^2 \phi 
\end{equation}
\begin{equation}
\label{A}
\Box \bar{h}_{33}= -\dfrac{4 \pi G \epsilon_g}{c^2}\, \partial_0^4 \,\bar{h}_{33}+\dfrac{2\pi G \epsilon_g}{c^2}\dfrac{\omega^2}{c^2k^2}\bigtriangleup\partial_0^2\,\bar{h}.
\end{equation}
Equation (\ref{newt}) results to be a modified Poisson equation for the macroscopic Newtonian potential: in \cite{17} it has been derived and analyzed in detail an analogous equation, discussing the peculiarity of the solutions both in strong and weak field. Now we calculate the $\left(0i\right)$ components of \eqref{FE}
\begin{equation}\label{0i}
\Box\bar{h}_{3i}=-\dfrac{4\pi G \epsilon_g}{c^2}\,\partial_0^4\,\bar{h}_{3i}-\dfrac{2\pi G \epsilon_g}{c^2}\dfrac{\omega}{ck}\,\partial_0^3 \partial_i \bar{h},
\end{equation}
together with the $\left(ij\right)$ components, resulting in
\begin{equation}\label{ij}
\Box \bar{h}_{ij}=-\dfrac{4\pi G \epsilon_g}{c^2}\,\partial_0^4\,\bar{h}_{ij}+\dfrac{2\pi G \epsilon_g}{c^2}\eta_{ij} \partial_0^4\, \bar{h}.
\end{equation}
Making use of \eqref{trace} we can compute a wave equation for the trace $\bar{h}$
\begin{equation}
\Box\bar{h}=\Box\bar{h}_{11}+\Box\bar{h}_{22}+\left(1-\dfrac{c^2k^2}{\omega^2}\right)\Box\bar{h}_{33}:
\end{equation}
$\Box\bar{h}_{11}$ and $\Box\bar{h}_{22}$ can be calculated from \eqref{ij}, whereas $\Box\bar{h}_{33}$ can be taken either from \eqref{A}, \eqref{0i} or even from \eqref{ij} itself. If one expresses $\Box\bar{h}_{33}$ from \eqref{A} obtains the following equation:
\begin{equation}
\Box \bar{h}= -\dfrac{2\pi G \epsilon_g}{c^2}\left(1-\dfrac{\omega^2}{c^2k^2}\right)\bigtriangleup \partial_0^2 \,\bar{h}.
\end{equation}
Making use of \eqref{0i} leads instead to
\begin{equation}
\Box \bar{h}= -\dfrac{2\pi G\epsilon_g}{c^2}\dfrac{\omega}{ck}\left(1-\dfrac{c^2k^2}{\omega^2}\right)\partial_0^3\,\partial_3\, \bar{h}.
\end{equation}
Lastly, from equation \eqref{ij} one gets
\begin{equation}
\Box\bar{h}=-\dfrac{2\pi G\epsilon_g}{c^2}\left(\dfrac{c^2k^2}{\omega^2}-1\right)\partial_0^4\,\bar{h}.
\end{equation}
Even if it seems a serious ambiguity that could outline a shortcoming of the procedure, when we investigate the real physics behind the three wave equations, \textit{i.e.} we compute the dispersion relation $\omega=\omega(k)$, we find that, in all three cases, the angular frequency satisfies
\begin{equation}
\omega(k)= \pm ck \quad ,\quad  \pm \sqrt{2}cm,
\end{equation}  
where the parameter $m^2=\dfrac{c^2}{4\pi G\epsilon_g}$ has been introduced. This means that $\bar{h}$ can be written as a superposition of a solution of d'Alembert equation, that can be canceled out with a further gauge transformation that preserves Hilbert gauge, plus an oscillation that does not propagate (the group velocity $v_g=d\omega/dk$ is zero) and has no observational meaning: we can erase the trace $\bar{h}$ bringing the count of degrees of freedom to five. At this stage all the components of the wave solve an identical wave equation:
\begin{equation}\label{waveeq}
	\Box\bar{h}_{\mu\nu}=-\dfrac{1}{m^2}\partial_0^4\,\bar{h}_{\mu\nu}.
\end{equation}	
Let us calculate the dispersion relation descending from equations \eqref{waveeq}:
\begin{equation}
\omega^2_{\pm}(k)=c^2 \bigg( -\dfrac{m^2}{2}  \pm \sqrt{\dfrac{m^4}{4}+m^2k^2} \bigg ) .
\end{equation}
It is immediate to verify that
\begin{equation}\label{omeghe}
\omega^2_+(k) \geq 0 \quad \forall\; k \quad \quad ,\quad  \quad  \omega^2_-(k) < 0 \quad \forall \, k.
\end{equation} 	
The fact that $\omega^2_-(k)$ is always negative implies that $\omega_-(k)$ are two purely imaginary solutions that characterize damped and growing modes.
However we will show that this branch of solutions does not contain any physical meaning. If we consider the solutions characterized by the plus sign and we perform the limit $\epsilon_g \to 0$, or $m^2 \to \infty$, \textit{i.e.} we remove the material medium, we find
\begin{equation}
\lim_{m^2 \to \infty} \omega^2_+(k)= c^2 k^2.
\end{equation}
This solution shows a good behavior: when we remove the material medium $\omega^2_+(k)$ goes back to be the dispersion relation of a vacuum solution. Performing the same limit on the solution characterized by the minus sign yields to
\begin{equation}
\lim_{m^2 \to \infty}\omega^2_-(k)=\lim_{m^2 \to \infty}-c^2m^2 \left(1+\frac{k^2}{m^2} \right)
\end{equation}
that is an infinite quantity. The fact that this solution does not go back to be a vacuum solution when we remove the material medium, performing the limit $\epsilon_g \to 0$, allows us to look at this branch of solutions as not physical. 
Then, the only solution that gives us physical information on the propagation of gravitational waves through a material medium is the one characterized by $\omega^2_+(k)$. As previously stated, $\omega^2_+(k)$ is always positive: this implies that the propagation is characterized by dispersion only. Let us calculate the group velocity:
\begin{equation}\label{groupvel}
v_g(k)=\dfrac{m^2 kc}{2\sqrt{\left(\sqrt{m^2k^2+\dfrac{m^4}{4}}-\dfrac{m^2}{2}\right)\left(m^2k^2+\dfrac{m^4}{4}\right)}}.
\end{equation}
We observe that the group velocity results always smaller than the speed of light.
Increasing the parameter $m$ causes the function $v_g(k)$ to reach small values in correspondence with increasing values of $k$; in the limit $m \to \infty$ the group velocity tends to be the constant $c$.
For a given $m$, if we consider $v_g(k)$ in the region $k\ll m$, we can approximate the function (\ref{groupvel}) with
\begin{equation}\label{vgapprox}
v_g(k) \simeq c \left(1-\dfrac{3k^2}{2m^2} \right)=c \left(1-\dfrac{6\pi G \epsilon_g}{c^2} k^2\right).
\end{equation} 
We will show that the condition \eqref{conditio} implies $k\ll m$ for all the realistic scenarios we will study. In this approximation the following equation holds
	\begin{equation}\label{omegaappross}
	\omega^2(k)=c^2k^2 \left( 1-\dfrac{k^2}{m^2}+O\left(\dfrac{k^4}{m^4}\right)\right).
	\end{equation}
The appearance of three extra degrees of freedom with respect to the vacuum case causes the fact that the wave possesses new longitudinal modes of oscillation. 
In order to analyze which kind of deformation is induced by each component on a sphere of test particles, we calculate the geodesic deviation equation, taken in the comoving frame
 \begin{equation}\label{testparticles}
\dfrac{1}{c^2}\dfrac{d^2\xi^\mu}{dt^2}=R^\mu_{\; \nu \rho \sigma}  u^\nu u^\rho \xi^\sigma=R^\mu_{\; 00i}  \xi^i,
 \end{equation}
 where $\xi^\mu=\left(0,\xi^x,\xi^y,\xi^z \right)$ is a vector denoting the separation between two nearby geodesics and the Riemann tensor is constructed up to the first order in $\bar{h}_{\mu\nu}$, as given in \eqref{metricamezzo}, together with the condition $\bar{h}=0$, that we exploit through the following gauge fixing:
 \begin{equation}
 \begin{split}
 \bar{h}_{11}&=\bar{h}_++\bar{h}^*\\
 \bar{h}_{22}&=-\bar{h}_++\bar{h}^*\\
\bar{h}_{33}&=\dfrac{2\omega^2}{c^2k^2-\omega^2}\bar{h}^*.
\end{split}
 \end{equation} 
  We fix the vector $\xi^\mu$ as
 \begin{equation}
\xi^\mu=\left(0,\xi^x_{(0)}+\delta\xi^x,\xi^y_{(0)}+\delta\xi^y,\xi^z_{(0)}+\delta\xi^z\right),
 \end{equation}
 being $\xi^x_{(0)}$, $\xi^y_{(0)}$, $\xi^z_{(0)}$ the initial positions and $\delta\xi^x$, $\delta\xi^y$, $\delta\xi^z$ the displacements of order $O(h)$ induced by the gravitational wave. We compute \eqref{testparticles} up to order $h$ separately for each component of $\bar{h}_{\mu\nu}$:
 \begin{enumerate}
 	\item [\RNum{1}]$\bar{h}_{+} \neq 0$ \\
 	\begin{equation}
 	\begin{split}
 	\dfrac{d^2 \delta \xi^x}{dt^2}&=-\dfrac{\omega^2}{2}  \bar{h}_{+} \xi^x_{(0)} \\
 	\dfrac{d^2 \delta \xi^y}{dt^2}&=\dfrac{\omega^2}{2}  \bar{h}_{+} \xi^y_{(0)} \\
 	\dfrac{d^2 \delta \xi^z}{dt^2}&=0.
 	\end{split}
 	\end{equation}
 	\item [\RNum{2}] $\bar{h}^* \neq 0$ \\
 	\begin{equation}
 	\begin{split}
 	\dfrac{d^2 \delta \xi^x}{dt^2}&=-\dfrac{\omega^2}{2}  \bar{h}^* \xi^x_{(0)}\\
 	\dfrac{d^2 \delta \xi^y}{dt^2}&=-\dfrac{\omega^2}{2}  \bar{h}^* \xi^y_{(0)}  \\
 	\dfrac{d^2 \delta \xi^z}{dt^2}&=-\left(c^2k^2-\omega^2\right)  \bar{h}^* \xi^z_{(0)}.
 	\end{split}
 	\end{equation}
 	\item [\RNum{3}] $\bar{h}_{12} \neq 0$ \\
 	\begin{equation}
 	\begin{split}
 	\dfrac{d^2 \delta \xi^x}{dt^2}&=-\dfrac{\omega^2}{2}\;\bar{h}_{12}\xi^y_{(0)}\qquad\qquad\qquad\\
 	\dfrac{d^2 \delta \xi^y}{dt^2}&=-\dfrac{\omega^2}{2}\;\bar{h}_{12}\xi^x_{(0)} \qquad\qquad\qquad\\
 	\dfrac{d^2 \delta \xi^z}{dt^2}&=0.
 	\end{split}
 	\end{equation}
 	\item [\RNum{4}] $\bar{h}_{13} \neq 0$ \\
 	\begin{equation}
 	\begin{split}
 	\dfrac{d^2 \delta \xi^x}{dt^2}&=\dfrac{c^2k^2-\omega^2}{2}\bar{h}_{13}\xi^z_{(0)}\\
 	\dfrac{d^2 \delta \xi^y}{dt^2}&=0 \\
 	\dfrac{d^2 \delta \xi^z}{dt^2}&=\dfrac{c^2k^2-\omega^2}{2}\bar{h}_{13}\xi^x_{(0)}.
 	\end{split}
 	\end{equation}
 		\item [\RNum{5}] $\bar{h}_{23} \neq 0$ \\
 	\begin{equation}
 	\begin{split}
 	\dfrac{d^2 \delta \xi^x}{dt^2}&=0\\
 	\dfrac{d^2 \delta \xi^y}{dt^2}&=\dfrac{c^2k^2-\omega^2}{2}\bar{h}_{23}\xi^z_{(0)} \\
 	\dfrac{d^2 \delta \xi^z}{dt^2}&=\dfrac{c^2k^2-\omega^2}{2}\bar{h}_{23}\xi^y_{(0)}.
 	\end{split}
 	\end{equation}
 \end{enumerate} 
The system \RNum{1} represents a standard plus polarization in the $xy$ plane. The deformation depicted by system \RNum{2} is a superposition of two independent modes: in the $xy$ plane we notice the presence of a breathing mode \cite{flavia,31,32} whilst along the $z$ direction is detectable a pure longitudinal stress in phase with the breathing, since $\omega^2 < c^2k^2$ for any $k$, with amplitude of the longitudinal deformation smaller than the amplitude of the breathing \footnote{In fact $c^2k^2-\omega^2<\omega^2$ is satisfied in the region $k<\sqrt{2}m$, but we will show that the condition \eqref{conditio} implies $k \ll m$.}. 
 The systems \RNum{3}, \RNum{4} and \RNum{5} describe cross polarizations in the $xy$, $xz$ and $yz$ plane respectively.
We can give an estimate of the ratio between the amplitudes of the anomalous polarizations and the standard one. We define the following quantities:
\begin{equation}
\begin{split}
\mathcal{A}^{\text{S}}&=\dfrac{\omega^2}{2}\\
\mathcal{A}^{\text{New}}&=\dfrac{c^2k^2-\omega^2}{2} ,
\end{split}
\end{equation}
where the superscript S stays for \textquotedblleft standard\textquotedblright. 
Making use of the approximate formula \eqref{omegaappross} we calculate the following estimate
\begin{equation}
\dfrac{\mathcal{A}^{\text{New}}}{\mathcal{A}^S}= \dfrac{k^2}{m^2}+O\left(\dfrac{k^4}{m^4}\right).
\end{equation}
In section \RNum{5} we will establish a pair of definite models of macroscopic medium in order to give a quantitative estimate of the ratio $k^2/m^2$, that acts as a precise marker of Macroscopic Gravity effects.
\section{Comparison with previous models} 
It is now necessary to point out some relevant differences between this work and Szekeres' model \cite{16}, given the fact that also in that work is described a derivation of a constitutive relation between the quadrupole tensor and vacuum fields.
We perturb the molecule with a vacuum gravitational wave expressed in TT gauge, \textit{i.e.} we act on the molecule uniquely with the two physical degrees of freedom possessed by the wave.
Szekeres perturbs the molecule with both the electric ($E_{\mu\nu}$) and magnetic ($B_{\mu\nu}$) part of Weyl tensor, then he discards the contribution arising from the magnetic part and lastly he performs the calculation in a frame in which the electric part of the Weyl is diagonal. This requirement, together with the trace free condition $E^\mu_{\;\mu}=0$, brings the number of degrees of freedom to two, but it can be shown that a vacuum gravitational wave that can be expressed as a diagonal tensor must be a purely plus polarized gravitational wave, \textit{i.e.} $E_{\mu\nu}$ cannot represent a generic vacuum perturbation. We argue that the nature of the vacuum fields chosen to perturb the molecule in Szekeres' work is not satisfactorily general. Another significant distinction is that Szekeres writes a constitutive relation for the induced quadrupole moment only, \textit{i.e.} he ignores the contribution of the static quadrupole\footnote{We stress that $T_{\mu\nu}^{(f)}$ does not contain any information about the static quadrupole of the molecules: it simply describes the energy momentum tensor of the centers of mass of the molecular dust.}. In our opinion this looks as a shortcoming of its work: in the unperturbed scenario the molecule is, on an averaged sense, spherical, but it possesses an intrinsic quadrupole proportional to the identity \eqref{quadrupintrinseco}, that must be reproduced through the second spatial derivatives of the Newtonian potential \eqref{matrice1}. This term comes to be source of a static contribution to the quadrupole. This difference between the two works produces a remarkable distinction when one calculates the Newtonian limit of the theory: in Szekeres case there is no modification on the Poisson equation for the Newtonian potential whereas we find a modified equation, \textit{i.e.} \eqref{newt}. The fact that the Newtonian potential inside the medium is modified in its shape by a global contribution, encoded in the value of $\epsilon_g$, coming from the whole medium, is physically reasonable. That being said, Szekeres' constitutive relation, when written in terms of a TT gauge gravitational wave , reads as
\begin{equation}
Q_{i0j0}=\epsilon_g \left( \dfrac{1}{4}h_{ij,00}+\dfrac{1}{4}\bigtriangleup h_{ij} \right).
\end{equation} 
It results identical to \eqref{matrela1} if one makes the assumption that Laplace operator acts on the wave in the same way the second time derivative does. This is certainly true in vacuum, but we stress the fact that the derivation of the constitutive relation is made inside the molecule. When we have to reproduce the components proportional to $\epsilon$ in equations \eqref{quadru}, we are forced to express it in terms of the second time derivative, because inside the molecule $\omega(k)$ is no longer equal to $ck$. 
As a consequence we obtain five dynamical degrees of freedom, both transverse and longitudinal, characterized by the same dispersion relation, whereas in Szekeres' model it can be shown that there are four transverse degrees of freedom, characterized by two different dispersion relation that coincide only in the short wavelength limit, and that the longitudinal degrees of freedom do not propagate. This cause a completely different phenomenology on test masses.
\section{Models of macroscopic medium}
Now we will give some quantitative estimates of $\epsilon_g$, without any intention of being too accurate, but merely realistic, in the characterization of the involved parameters. Firstly we simplify the expression \eqref{epsilone} for the gravitational dielectric constant $\epsilon_g$: remembering that $\omega_0^2=\frac{4}{3}\pi G\rho_0$ and also that the molecule is described as a sphere with constant mass density $\rho_0$ we get to the following simplified expression
\begin{equation}
\epsilon_g=\dfrac{NL^5c^2}{4G}.
\end{equation} 
This means that, in order to characterize a macroscopic medium composed by molecule that are spheres with constant mass density, it is necessary to assign only the values of the radius of the molecule and the density of molecules: the dielectric response of the medium is independent from the mass of the molecules. We will describe two different models of the material medium: the first assuming binary systems to be the molecules of the medium, the second will be composed by open clusters.
Let us begin with the first case: the molecules are binary systems.
It was widely believed \cite{33,34,35,36} that the major part of the stars in our galaxy was part of binary or multiple systems: more recent observations changed this paradigm. A better estimate \cite{37} is that approximately one third of the stars in our galaxy are located in binary, or with higher multiplicity, systems; the remaining two thirds can be safely considered single stars. We will calculate the dielectric constant in three different regions of the Galaxy, from the edge to the center of it: $(i)$ is a region characterized by a stellar density of $1 \, pc^{-3}$, in $(ii)$ this value increases to $100 \, pc^{-3}$ and in $(iii)$ it reaches $10^5 \, pc^{-3}$. We will set $N$ as one third of these values in each region. The size of a binary system is a quite variable parameter: it can go from less than one to some thousands of AU. However, the typical distances are those of the Solar system, rather than the typical distances between stars (few light years)\cite{36}: we will set the radius of the molecule to $100 \; AU=1.5 \cdot 10^{13} \, m$. We get the following three values for the gravitational dielectric constant:
\begin{itemize}
	\item [$(i)$]$\epsilon_g=2.87 \cdot 10^{42} \, kg \; m \qquad \quad m^2=3.74 \cdot 10^{-17} \, m^{-2}$ \\
	\item [$(ii)$]$\epsilon_g=2.87 \cdot 10^{44} \, kg \; m \qquad \quad m^2=3.74 \cdot 10^{-19} \, m^{-2}$\\
	\item[$(iii)$]$\epsilon_g=2.87 \cdot 10^{47} \, kg \; m \qquad \quad m^2=3.74 \cdot 10^{-22} \, m^{-2}$.
\end{itemize}
The size of the molecule has been set to $L= 1.5 \cdot 10^{13} \; m$. Hence, only gravitational waves with wavenumber $k \ll 1/L =6.68 \cdot 10^{-14} \; m^{-1}$ satisfy the condition \eqref{conditio}. We consider gravitational radiation with wavenumber $k=6.68 \cdot 10^{-16} \, m^{-1}=10^{-2} \frac{1}{L}$.
It is easy to check that $\frac{1}{L}<m$; hence the condition $k\ll m$ is satisfied and we are allowed to use formula (\ref{vgapprox}) to calculate the group velocity of the wave inside the medium.
If we take the maximum value obtained for $\epsilon_g$ $(iii)$ we find
\begin{equation}
v_g(k)=c \left (1-1.79 \cdot 10^{-9}\right).
\end{equation}
Taking for $\epsilon_g$ the value $(ii)$ yields to
\begin{equation}
v_g(k)=c \left (1-1.79 \cdot 10^{-12}\right).
\end{equation}
If we consider for $\epsilon_g$ the minimum value $(i)$, we find
\begin{equation}
v_g(k)=c \left (1-1.79 \cdot 10^{-14}\right).
\end{equation}
With the chosen values of the parameters we observe that the ratio $k^2/m^2$ is at most $10^{-9}$: as we have shown, this quantity indicates the ratio between the amplitudes of the anomalous polarizations and the standard one.
Actually binary system can only be roughly approximated via the model of a spherical molecule, since they posses an intrinsic quadrupole. However,
the random orientation of the orbital planes of these
binaries with respect to the direction of propagation of the
incoming gravitational wave, leads us to infer that the
average effect can be qualitatively estimated even in the
present simplified framework. An improvement of the model is currently under development.
 
Now we consider the case of a material medium whose molecules are open clusters \cite{38,39}.
Our galaxy is estimated to contain about $100,000$ open clusters, mainly located in the central disc. We will calculate the density of molecules as the number of molecules ($10^5$) divided by the volume of the Galaxy, assuming it to be a disc with diameter of $30 \, kpc$ and thickness of the disc of about $0.6 \, kpc$ (roughly the size of the Milky Way \cite{40}). For the density of molecules we calculate the value $N=8.03 \cdot 10^{-57} \, m^{-3}$. The radius of the molecule will be fixed to $L=3 \; pc=9.26 \cdot 10^{16} \, m$. Under these assumptions we get the following value for the gravitational dielectric constant:
\begin{equation}
\epsilon_g=1.84 \cdot 10^{55} \, kg \, m \qquad \quad m^2=5.84 \cdot 10^{-30}\, m^{-2}.
\end{equation}
We consider gravitational radiation with wavenumber $k=10^{-2} \frac{1}{L}=1.08 \cdot 10^{-19} \, m^{-1}$: once again the condition $k \ll m$ is satisfied and we use the equation (\ref{vgapprox}) to calculate the dispersion. We obtain
\begin{equation}
v_g(k)=c \left (1-3.01 \cdot 10^{-9}\right).
\end{equation}
As in the case of the first model, the ratio $k^2/m^2$ is roughly of order $10^{-9}$. 

We focused our analysis on two specific regions of wavelengths. In the first case (binaries) the signal is characterized by $\lambdabar \simeq 10^{15} \, m$, that is comparable with the scale of lengths to which the space interferometer LISA will be sensitive \cite{41,42}. In the second case (clusters) the wavelength of the signal is in the region $\lambdabar \simeq 10^{18} \, m$ and it is, in principle, inside the sensitivity curve of experiments like IPTA \cite{43,44}.

Our analysis demonstrates that, both from a conceptual and
phenomenological point of view, the deformation that a
gravitational wave induces on the bounded systems in the
Galaxies effectively modifies the nature of the wave itself, generating longitudinal polarization
modes and, over all, a subluminal velocity of propagation, in contrast to the vacuum case.
Despite the deviation from the speed of light is
at most of order $10^{-9}$, its integrated
effect over very large distances can produce a significant
delay with respect to the electromagnetic signal.
This scenario is expected of significant impact in
the analysis of the ``follow up'' of astrophysical
sources and this feature must be taken into account
in the set up of a gravitational astronomy.

\section{Conclusions}

The main merit of the analysis above is to give 
significant phenomenological implications when the case of a gravitational wave crossing a medium characterized by a molecular structure is considered.
We revise the original and valuable approach presented in \cite{16} 
only in avoiding the use of Weyl tensor components as 
basic gravitational field variables. 
We simply adopt the space-time ripple as the natural variable 
set to describe the gravitational field both in 
vacuum and within the matter.
In fact, the vacuum gravitational wave modifies, via the geodesic deviation
equation, the molecule morphology, inducing in this manner an effective (average) quadrupole 
contribution, which is calculated just via 
the resulting displacements inside the 
molecule. More specifically, we 
calculate an expression of the induced quadrupole components in terms of the second time derivatives of the vacuum gravitational wave amplitudes (whose spatial variation is
neglected when calculating the geodesic deviation).
Furthermore, we also account for the deformation of the 
Newtonian potential living within the self-gravitating medium. 
In fact, we reproduce through the second spatial derivatives of the Newtonian potential the non-zero quadrupole tensor 
also for the unperturbed molecule and this causes the emergence of a macroscopic gravity effect 
also on the static level. Poisson equation acquires, 
as result of the medium, a biharmonic term, 
whose relative amplitude with respect to the standard Laplacian term 
is fixed by the same macroscopic gravity parameter that modifies the 
gravitational wave propagation.
This result is in close analogy to the one obtained in \cite{17}, where an \textit{ad hoc} (strong field) hypothesis on the link existing between the 
quadrupole tensor and the Riemann tensor is postulated.
The emergence of a net
non-zero quadrupole in the field equation modifies the wave propagation, introducing new
effective modes of oscillation and implying a subluminal speed of these 
ripples. 
The new polarization modes and the 
group velocity we obtain are a precise marker of the present approach and 
offer a phenomenological tool to 
search their signature in data analyses of incoming experiments, like LISA or IPTA \cite{41,42,43,44}. 
However, it is relevant to stress that, 
in the limit of large scales (small wavenumbers) of the wave, the group
velocity 
expression overlaps the corresponding limit in the analysis developed in 
\cite{16}. 
Such a degeneracy of the two approaches 
when implementation of the model is 
performed for real and relevant astrophysical systems, can be regarded as a
reciprocal validation of the two approaches 
on a physical level: 
when the modifications concern large scales with respect to the characteristic
length scale of the macroscopic gravity effect, different representations of 
the constitutive relation provides the same weak features on the gravitational
wave propagation. 
In \cite{16} the use of Weyl tensor was mainly due to the construction of a parallelism with the electromagnetic case, in the spirit of constructing a "gravitational induction'' tensor, which however can not be consistently defined along such a parallelism. Also our approach does not provide an induction tensor for the gravitational interaction, but it relies on the metric perturbation as the only quantity to be treated like an independent one in the procedure. Such a natural choice has important implications on the exact form of the constitutive relation and of the dispersion relation. 
	We get different polarization modes with respect to the ones outlined in \cite{16} and the gravitational ripples have, within the gravitational medium, five independent degrees of freedom. 		
	It remains, as an open issue, the construction of a 
	suitable gravitational induction field, based on the 
	parallelism with the electromagnetism with matter, but 
	restricting the study to the weak field only. In fact, the valuable splitting of the total energy-momentum tensor of the medium into a free contribution  (the molecule 
	center of mass) and a quadrupole tensor is valid 
	only for small deviation of the particle trajectories from the unperturbed scenario.

\vspace{0.8cm}
\begin{acknowledgments}
FM would like to thank Flavio Bombacigno for the fruitful discussions regarding the interpretation of the results.
\end{acknowledgments}

\end{document}